\documentclass{aa}
\usepackage{amsmath,amssymb,booktabs}
\usepackage{graphicx}
\usepackage{txfonts}
\usepackage{placeins}
\usepackage{xcolor}
\usepackage{stfloats}
\usepackage[
    colorlinks=true,
    citecolor=blue,
    linkcolor=black,
    urlcolor=blue
]{hyperref}
\bibpunct{(}{)}{;}{a}{}{,}
\newcommand{\Lmax}{L_{\max}}

\newcommand{\Tact}{T_{\rm act}}
\newcommand{\ClassI}{Class~I}
\newcommand{\ClassII}{Class~II}
\titlerunning{Repeater Activity Spans with Cluster Lensing}
\authorrunning{H. Zhou et al.}

\newcommand{\GeometryCeiling}{79.29}

\begin{document}
\title{Probing the Activity History of Repeating Fast Radio Bursts with Cluster Lensing}
\author{Huan Zhou\inst{1}
\and Kai Liao\inst{2}
\and Zhengxiang Li\inst{3,4}
\and Cheng-Gang Shao\inst{1}}
\institute{School of Physics and Optoelectronic Engineering, Yangtze University, Jingzhou 434023, China
\and School of Physics, Wuhan University, Wuhan 430072, China\\
\email{liaokai@whu.edu.cn}
\and School of Physics and Astronomy, Beijing Normal University, Beijing 100875, China\\
\email{zxli918@bnu.edu.cn}
\and Institute for Frontiers in Astronomy and Astrophysics, Beijing Normal University, Beijing 102206, China}
\date{}

\abstract{
Fast radio bursts (FRBs) are millisecond duration coherent radio transients, some of whose sources repeat, but the nature and active lifetimes of their engines remain uncertain. Strong lensing by galaxy clusters provides a physical lever on this lifetime problem: a short campaign can sample emission epochs of a repeating FRB separated by years, and distinct bursts in different lensed images, given a known relative lens delay, set a lower bound on its active lifetime, whereas delayed copies of a single burst only encode the lens delay. Based on the cluster-lens model of MACS J0717.5+3745, we quantify the accessible range of emission epochs for finite activity episodes, observing windows, and discovery selection. For a source-frame episode of duration \(100~{\rm yr}\) at \(z_s=2\) and a source-frame rate of \(0.1\,{\rm h}_{\rm src}^{-1}\), a five year campaign conditional on discovery yields a \(95\%\) gain threshold of \(m_{95}\simeq25.6\), corresponding to a source-frame span of \(L_{95}\simeq42.6~{\rm yr}\). At the population level, combining distinct burst associations with same burst echoes provides a way to test specified activity durations with a finite lens catalogue, e.g., activity durations differing by an order of magnitude can be distinguished with expected Bayes factor above $20$. More broadly, these lensed source frame epochs enable applications beyond activity-duration tests, e.g., linking burst activity to the evolution of the engine and its environment. Galaxy-cluster lensing therefore provides a promising future probe for accessing otherwise unobserved source history and for distinguishing between different activity duration models.
}
\maketitle
\nolinenumbers
\raggedbottom

\section{Introduction}\label{sec:intro}
Fast radio bursts (FRBs) are millisecond duration, coherent radio transients that are mostly extragalactic~\citep{Lorimer2007,Thornton2013}. Observationally, FRB sources are classified into repeaters and apparently non-repeating sources~\citep{Petroff2019,Petroff2022}. However, because this classification depends on observing exposure and sensitivity, it need not imply that repeaters and apparently non-repeating sources constitute two physically distinct populations~\citep{James2023,Yamasaki2024,Kirsten2024}. The repeater FRB~121102 demonstrated that at least some FRB engines survive individual burst-producing events and enabled a direct host-galaxy association~\citep{Spitler2016,Chatterjee2017}. Repeaters often show intermittent activity, characterized by clustered waiting times, periodic visibility windows, and time-variable burst rates~\citep{Opperman2018,CHIME2020,Li2021,Abbott:2026jfn}. Consequently, nondetections depend on sensitivity, cadence, instantaneous burst rate, and activity history, and therefore do not imply that the source has ceased~\citep{Kumar2025}. The duration of activity is distinct from the emission mechanism, e.g., Galactic magnetar bursts support a magnetar origin for at least some FRBs~\citep{CHIMEMagnetar2020, Bochenek2020}, whereas the globular-cluster repeater indicates that repeating sources can occur in old stellar environments~\citep{Kirsten2022}. Candidate engines include magnetically powered neutron stars, energetic pulsars, and accreting compact objects~\citep{CordesWasserman2016,Beloborodov2017,Platts2019,Sridhar2021,Zhang2023}, whose observable activity may depend on the energy-reservoir evolution, external nebular conditions, or mass transfer~\citep{Metzger2017,Metzger2019,MargalitMetzger2018,SridharMetzger2022}. Constraining the duration of a specified active phase is therefore physically informative even if different engines can produce similar burst properties. Direct monitoring is limited in two respects, i.e., intermittent activity makes quiescent intervals an ambiguous signature of cessation~\citep{Opperman2018,CHIME2020,Li2021,Abbott:2026jfn,Cook2026}, and the observed repeater fraction depends on sensitivity and accumulated exposure~\citep{CHIME2021,Andersen2023,James2023,Yamasaki2024, Kumar2025,Cook2026}. A higher cadence provides denser sampling within the accessible interval, but it cannot recover emission from epochs earlier than that interval.

Strong gravitational lensing supplies a distinct channel of repeated FRB signals: a single burst can traverse multiple paths through a foreground gravitational potential and arrive as spatially separated images at different times, whereas an intrinsically repeating source produces image streams that share a common event sequence but are shifted by the lens delays. Due to their short intrinsic timescales, high event rates, and extragalactic origin, lensed FRBs serve as precision probes of astrophysics and cosmology. Lensing signatures span a wide range of time delays across different lens masses, e.g., spectral distortions~\citep{Kader2022,Leung2022} and short-delay echoes~\citep{Munoz2016,ConnorRavi2023,Zhou2026} probing compact dark matter with masses below those of intermediate-mass black holes; image time delays from galaxy-scale lensing enabling measurements of \(H_0\)~\citep{Li2018,Wucknitz:2020spz,Zhang:2024rra}; timing of resolved lensed images constraining the motion of the emission region~\citep{DaiLu2017}; and equivalence-principle tests~\citep{Reischke2022,Jiang2024}. Galaxy clusters provide a rich lensing geometry with multiple images and widely separated image time delays~\citep{KneibNatarajan2011}, linking transient lensing to source physics and cosmography~\citep{Oguri2019,Natarajan2024}. For an intrinsically repeating FRB, this geometry produces image streams that share a common event sequence but are shifted by the lens delays, so different images can sample different emission epochs. Previous galaxy cluster-lensed FRB forecasts have focused on detection rates, image multiplicity, and cluster reconstruction~\citep{Wagner2019}. Beyond these forecast-level questions, the wide delays of cluster lenses also make them suitable for constraining the active lifetime of repeating FRBs. 

In this paper, we investigate how galaxy-cluster lensing extends the source history accessible to a finite monitoring campaign. A~\ClassI\ pair comprises distinct intrinsic bursts from the same source detected in different images, whereas a~\ClassII\ pair repeats a single emission epoch; for a~\ClassI\ pair, subtracting the known lens delay from the arrival-time difference yields the source-frame separation between the two bursts, thereby constraining the activity span of repeating FRBs. Based on the lens reconstructions of MACS J0717.5+3745~\citep{Johnson2014,Limousin2016,Lotz2017,HFFArchive}, we calculate the gain in the maximum detectable activity duration and in the number of distinct sampled bursts. We further propose the joint occurrence of the two classes to test hypothesized activity durations. This joint statistic can serve as a new probe of repeating FRB models, and we discuss its future applications in astrophysics. This paper is organized as follows. Section~\ref{sec2} describes the cluster-lens model, source activity, and observing selection. Section~\ref{sec3} defines the activity-span and burst-count observables and develops the joint likelihood for the two classes. Section~\ref{sec4} presents predictions for activity-duration constraints and astrophysical applications. Section~\ref{sec5} presents our conclusions.

\begin{figure*}[!htbp]
\centering\includegraphics[width=0.9\textwidth]{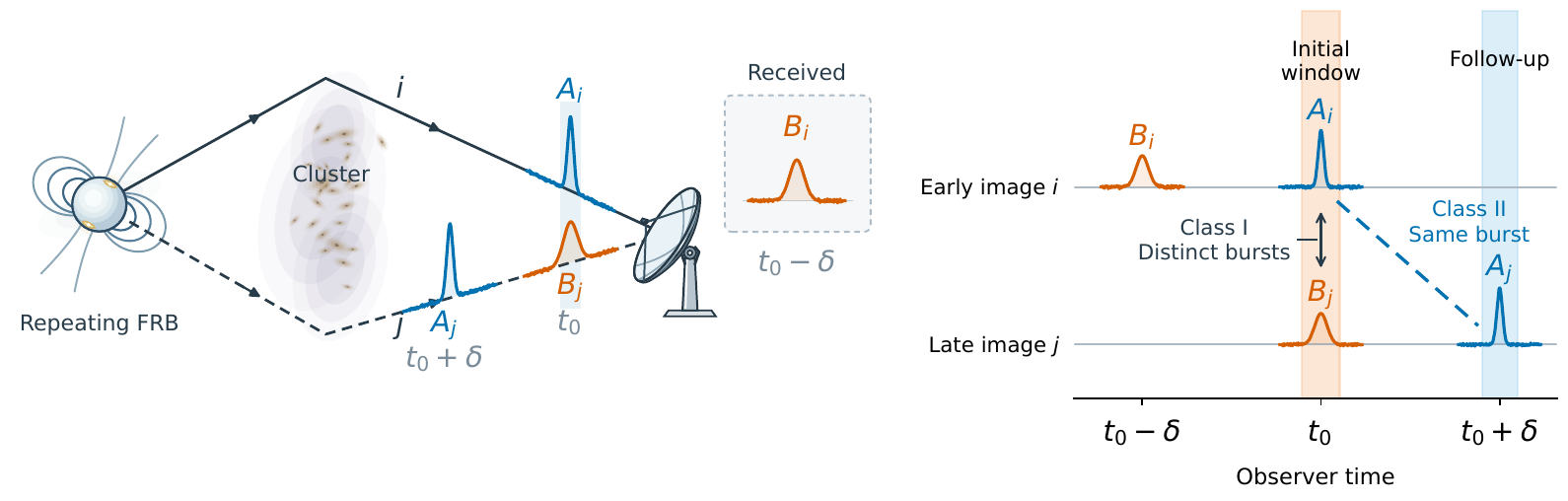}
\caption{Schematic of a cluster-lensed repeating FRB.~\textbf{Left Panel:} A repeating FRB, the cluster, and a radio telescope are connected by two representative image paths. $B_i$ arrives at $t_0-\delta$, $A_i$ and $B_j$ coincide at $t_0$, and $A_j$ is expected at $t_0+\delta$.~\textbf{Right Panel:} The vertical $A_i-B_j$ association denotes the coincident-arrival special case of Class~I, while the blue $A_i-A_j$ connection denotes Class~II.}\label{fig:schematic}
\end{figure*}

\section{Cluster-Lensed Repeating FRBs}\label{sec2}
\subsection{Cluster Lensing Model}
For a cluster lens model with lensing potential \(\psi\), the image positions \(\boldsymbol{\theta}_i\) of a source with angular position \(\boldsymbol{\beta}\) and redshift \(z_s\) satisfy the lens equation. We adopt the standard thin-lens arrival-time convention~\citep{TreuMarshall2016}, in which the arrival-time surface is given by
\begin{equation}
t(\boldsymbol\theta)=\frac{(1+z_l)D_lD_{s}}{cD_{ls}}
\left[\frac{|\boldsymbol\theta-\boldsymbol\beta|^2}{2}-\psi(\boldsymbol\theta)\right],
\label{eq:fermat}
\end{equation}
where $z_l$ is the lens redshift and $c$ is the speed of light. The time delay distance $D_{\Delta t}\equiv(1+z_l)D_lD_s/D_{ls}$ has dimensions of length ($D_l$, $D_s$, and $D_{ls}$ are the angular-diameter distances from observer to lens, observer to source, and lens to source, respectively), and the dimensionless Fermat potential is $\phi(\boldsymbol{\theta},\boldsymbol{\beta})=|\boldsymbol{\theta}-\boldsymbol{\beta}|^2/2-\psi(\boldsymbol{\theta})$. For a fixed source position, the pairwise time delay $d_{ji}$ is
\begin{equation}
 d_{ji}\equiv\Delta_j-\Delta_i=\frac{D_{\Delta t}}{c}\left[\phi(\boldsymbol{\theta}_j)-\phi(\boldsymbol{\theta}_i)\right].
 \label{eq2-1-2}
\end{equation}
Figure~\ref{fig:schematic} illustrates the two classes of lensing signatures considered in this work. A repeating FRB source emits several intrinsic bursts that propagate along different gravitationally lensed image paths. In Class~I, signals detected in different image streams correspond to distinct intrinsic bursts, such as \(A_i\) and \(B_j\), whose arrival times can overlap in time because each image stream carries a different lens delay. In Class~II, the signals are multiple delayed images of the same intrinsic burst, such as \(A_i\) and \(A_j\), i.e., one image arrives first, and its delayed counterpart appears later at the predicted image position. The two classes therefore probe different aspects of the source: Class~I constrains the source activity span between two distinct bursts, whereas Class~II tests the detectability of multiple lensed images of a single burst. Defining $\Delta_i=t(\boldsymbol\theta_i)-\min_jt(\boldsymbol\theta_j)$ as the observer-frame image delay relative to the earliest image ($\Delta_1=0$), the observer-frame arrival time of intrinsic burst \(k\) in image \(i\) is then
\begin{equation}
t_{ik}=(1+z_s)s_k+\Delta_i,\qquad
\label{eq:mapping}
\end{equation}
where $z_s$ is the source redshift and the factor $(1+z_s)$ accounts for cosmological time dilation between the source and the observer. We employ the 201 released CATS v4.1 range maps of MACS J0717.5+3745, an HFF cluster in which multiply imaged galaxies constrain a complex mass distribution~\citep{Johnson2014,Limousin2016,Lotz2017,HFFArchive}. We treat the released potential as an effective single-plane potential normalized to \(D_{ls}/D_s=1\) and adopt its effective plane redshift \(z_l=0.54\) consistently in the distance scaling and time delay distance, together with a flat cosmology with \(H_0=70\ \mathrm{km\,s^{-1}\,Mpc^{-1}}\) and \(\Omega_{\rm m0}=0.3\). We consider 201 maps and generate 128 source position proposals per map, yielding 25728 proposals in total. All maps are assigned equal prior weight, and each source proposal carries an area weight proportional to $A_{\mathrm{prop},m}/128$ (the sampling region is constructed from the source-plane caustics of the lens model at a given source redshift, since these caustics mark where the image multiplicity changes and thereby delineate the potential multiply imaged source positions.).

\begin{figure*}[!htbp]
\centering\includegraphics[width=0.9\textwidth]{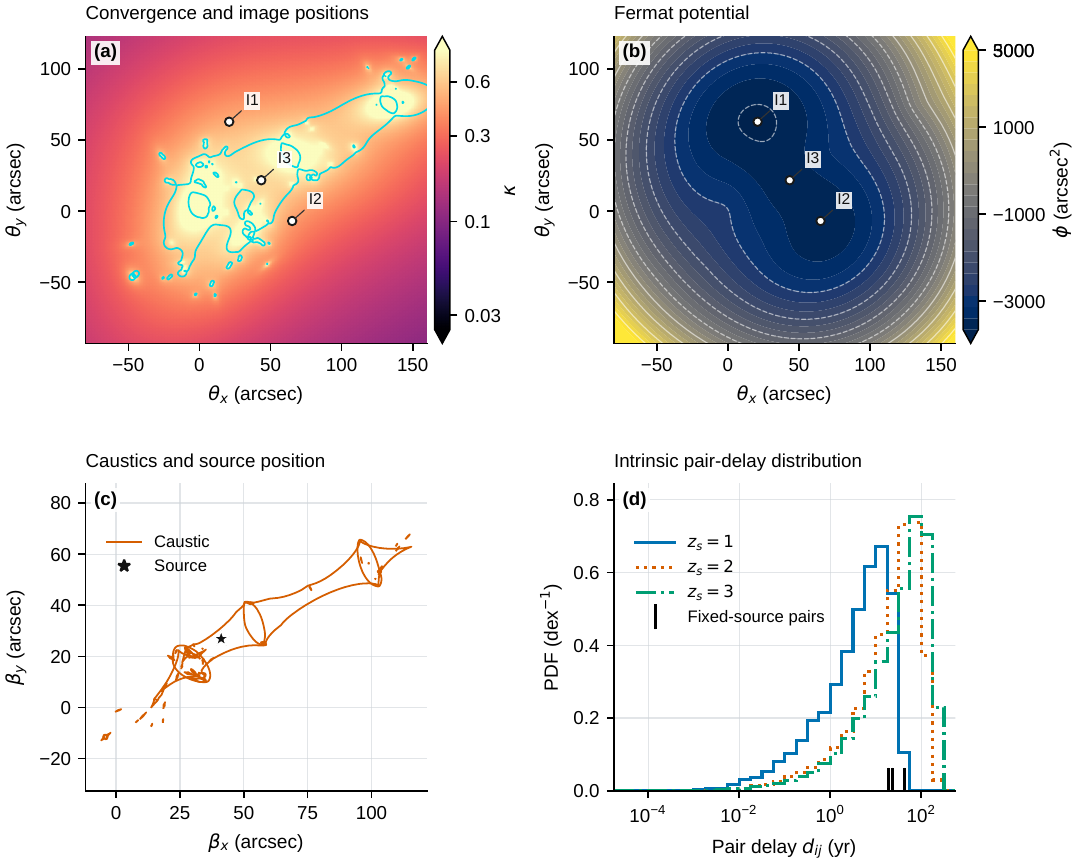}
\caption{MACS~J0717 lensing calculation based on the HFF CATS v4.1 range maps. Panels (a)--(c) show the convergence map, Fermat potential surface, and source-plane caustics for range map 000 with a representative source at $z_s=2$ with position $(41.17,26.89)''$. Panel (d) shows the normalized probability distributions of pair-delay for $z_s=[1, 2, 3]$, constructed from the recovered image pairs in the 201-map ensemble, using all source positions at each redshift.}\label{fig:cluster}
\end{figure*}

Figure~\ref{fig:cluster} shows the simulated geometry and the distribution of time delay before imposing a monitoring window. Panels~(a)--(c) show a representative convergence map, Fermat-potential surface, and source-plane caustic structure for MACS~J0717 with a source at $z_s=2$ and position $(41.17,26.89)''$. Panel~(d) presents the probability distributions of the observer-frame time delays between all image pairs at the three source redshifts $z_s=[1, 2, 3]$. As a specific example at $z_s=2$, solving the lens equation for the source position above yields three images with observer-frame delays $(0,23.22,42.01)\ \mathrm{yr}$, corresponding to pair delays of $18.79$, $23.22$, and $42.01\ \mathrm{yr}$ as marked in panel~(d). For the five-year observing calendar, the 42.01 yr time delay between the first and third images extends the geometrical reach in source-frame time from 1.67 yr for a single image to 15.67 yr. The lensing advantage is therefore access to a longer baseline in the source's activity history. The largest time delay separation between distinct bursts detected in different images provides constaints on the activity span.

\subsection{Source Activity and Observing Selection of Repeating FRBs}
At observing band for repeating FRBs, we adopt an unlensed burst-fluence distribution with a complementary cumulative form, \({\rm CCDF}(F)=(F/F_{\min})^{-\gamma}\) for \(F\geq F_{\min}\). For a sharp detection threshold \(F_{\rm lim}\), an image is detectable when \(|\mu|F\geq F_{\rm lim}\), yielding the single-image detection efficiency
\begin{equation}
S(\mu)=\min[1,(|\mu|/\mu_\ast)^\gamma],
\label{eq:pairweight}
\end{equation}
where \(\mu_\ast=F_{\rm lim}/F_{\min}\), and we take the choices \(\gamma=1/2\)~\footnote{The choice $\gamma=1/2$ is motivated by the shallow cumulative high-energy burst distribution measured for FRB 20201124A~\citep{Kirsten2024}, with a comparable flattening reported for FRB 20220912A~\citep{OuldBoukattine2026}.} and \(\mu_\ast=10\). For an observer frame window, the detection indicator for image \(i\)
\begin{equation}
Y_i(s,u)=W_i((1+z_s)s+\Delta_i)\mathbf1_{u<S_i}
\label{eqci}
\end{equation}
where \(\mathbf{1}_{u<S_i}\) is an indicator function that equals one if \(u<S_i\) and zero otherwise, $u\sim U(0,1)$ is drawn independently for each intrinsic burst to assign its unlensed fluence and is shared by all lensed images of that burst, $W_i(t)$ denotes its representation in observer time. For non-overlapping visits with start and end times $v_a^-$ and $v_a^+$, the observing window $W(t)$ in observing frame  is
\begin{equation}
W(t)=\sum_a\mathbf{1}_{[v_a^-,v_a^+)}(t),
\label{eq:visits}
\end{equation}
where the indicator selects the visit $a$, and its observer-frame interval $[v_a^-,v_a^+)$ maps to the source-frame interval $[(v_a^--\Delta_i)/(1+z_s),(v_a^+-\Delta_i)/(1+z_s))$. We adopt an exposure-equivalent daily transit model.~\citet{CHIME2020} reported 64 usable hours on FRB~20180916B between 2018 August 28 and 2019 September 30, obtained by summing daily beam transits and excluding interruptions. This corresponds to an average of approximately ten minutes per day. We approximate this exposure with ten-minute windows centered at \(t_j=(j+1/2)P_{\rm sid}\), where \(P_{\rm sid}=86164\) s, within \(0<t<T_{\rm obs}=5\) yr, yielding 1831 windows and a total exposure of 305.17 hr. We then define \(W_i(s)\) as the observing window function for image \(i\) in source-frame time, which indicates whether emission at source-frame time \(s\) falls within the observing window of image \(i\) under the above observing strategy.

For a given repeating FRB model, the rate entering the image-detection calculation is modeled as constant activity within a single source-frame episode \([b,b+\Tact)\)
\begin{equation}
r(s)=\lambda_0\,\mathbf{1}_{[b,b+\Tact)}(s),
\label{eq:rate}
\end{equation}
where $b$ is the source-frame time of the activity onset measured relative to the start of the monitoring campaign. We compare Poisson burst rates of $\lambda_0=0.1~{\rm h}_{\rm src}^{-1}$ and $0.01~{\rm h}_{\rm src}^{-1}$, corresponding to mean source-frame waits of 10 and 100 hr before image-dependent selection. Magnetar-flare shocks provide a useful reference timescale for the engine: the model of~\citet{Metzger2019} predicts major ion-rich ejections separated by \(\Delta T\sim10^5~{\rm s}\), i.e., \(\Delta T^{-1}\simeq0.036~{\rm h}_{\rm src}^{-1}\). Magnetospheric emission~\citep{LuKumarZhang2020} and flaring-binary scenarios~\citep{Sridhar2021} likewise motivate recurrent activity. Since the predicted burst rates in these scenarios depend on source properties and emission-selection criteria, we treat \(\lambda_0\) as a shared phenomenological rate parameter.

\section{Activity Span and Distinct-Burst Counts for Cluster-Lensed Repeating FRBs}\label{sec3}
\subsection{Activity-Span and Burst-Count Gains from Cluster-Lensed Images}\label{sec3-1}
For distinct bursts $k\ne l$ observed in different images, their source epoch separation is
\begin{equation}
L_{ijkl}=\frac{|(t_{jl}-t_{ik})-(\Delta_j-\Delta_i)|}{1+z_s}=|s_l-s_k|,
\label{eq:span}
\end{equation}
where \(i\) and \(j\) index images, while \(k\) and \(l\) index intrinsic bursts rather than image copies. For \(k=l\), the arrival-time difference between two images is solely the signed lens delay, so the delay-corrected separation is zero regardless of the observed arrival-time separation. For \(k\neq l\), subtracting the signed lens delay and dividing by \(1+z_s\) recovers the source-frame emission interval. We retain a single maximum per lensing source  as
\begin{equation}
  \Lmax=\max_{\substack{i<j,\ k\ne l\\\text{detected pairs}}} L_{ijkl}.
  \label{eq:max}
\end{equation}
We set \(L_{\max}=0\) when no pair of distinct bursts is detected in different images. Otherwise, provided that all detected bursts belong to a single activity episode, we have \(T_{\rm act}\geq L_{\max}\). To quantify the additional source-history leverage provided by lensing relative to the monitoring duration \(T_{\rm obs}\), we define
\begin{equation}
G \equiv \frac{(1+z_s)\Lmax}{T_{\rm obs}},
\label{eq:gain}
\end{equation}
A single image yields $G\leq1$, whereas delayed images can yield $G>1$ by sampling source epochs substantially earlier or later within monitoring duration. For a gain threshold $m$, the equivalent source frame threshold is
\begin{equation}
  \ell_m \equiv \frac{mT_{\rm obs}}{1+z_s},
  \qquad
  G>m \;\Longleftrightarrow\; \Lmax>\ell_m.
  \label{eq:gain-threshold}
\end{equation}
The probabilities for \(m=5\), \(10\), and \(20\) correspond to observer-frame equivalent baselines of 25, 50, and 100 yr, respectively.

Assuming a constant activity-onset rate, we draw the source-frame onset epoch \(b\) uniformly over \([-T_{\rm act},\,T_{\rm obs}/(1+z_s)]\), conditional on the activity episode overlapping the monitoring window. This range contains all onset epochs for which the episode can overlap the reference-image monitoring window. Over this range, the finite number of visits and the flux threshold set the probability that the episode is included in the sample. Let $a_{\rm ref}(s)=S_1W_1(s)$ and $a_{\rm union}(s)=\max_i[S_iW_i(s)]$ denote the detection efficiency at source frame time $s$ for a burst in the reference image and in at least one image, respectively. For $X={\rm ref},~{\rm union}$, the number of distinct intrinsic bursts detected in case $X$ is~\footnote{We adopt a reference-image-triggered search: absent an independently established association between image streams, a burst detected in the minimum-delay image initiates archival and follow-up searches. Lens-model constraints guide the search for earlier intrinsic bursts and schedule observations at the predicted arrival times of other images. A reference-image detection defines the sample considered here, although detections in other images could also trigger such searches.}
\begin{equation}
\mu_X(b,g,\Tact)=\nu_0\int_b^{b+\Tact}a_X(s)\,ds.
\label{eq:countmean}
\end{equation}
where $g$ denotes a mass-map/source-position configuration with weight $w_g=A_{{\rm prop},m}/(201\times128)$, and $\nu_0=8766\lambda_0$ is the rate per source year. The discovery probability of at least one burst in the reference image during the observer frame year is
\begin{equation}
p_{\rm ref}(b,g,\Tact)= 1-\exp\!\left[-\mu_{\rm obs}(b,g,\Tact)\right].
\label{eq:onset}
\end{equation}
Our forecast for the Class~I association probability above a gain threshold $m$ is
\begin{equation}
\begin{split}
P(G>m, \Tact)=\frac{\sum_g w_g P(G>m, g, \Tact)}
{\sum_g w_g P(g,\Tact)} ,
\end{split}
\label{eq:Q}
\end{equation}
where $P(g,\Tact)=\langle p_{\rm ref}(b,g,\Tact)\rangle_b$ is the onset marginalized probability, $P(G>m, g, \Tact)$ is the geometry-weighted probability of detecting at least one reference-image burst and a Class~I pair with \(G>m\) marginalized over onset times. We quantify the increase in the number of distinct intrinsic bursts detected when all lensed images are used instead of the reference image alone. For sources with at least one reference-image detection, the conditional expectations and their ratio are
\begin{equation}
\begin{aligned}
&\mathbb E[N_X\mid N_{\rm ref}\geq1]
=\frac{
\left\langle
p_{\rm ref}\mu_X+
\mu_{\rm ref}e^{-\mu_{\rm ref}}
\right\rangle_{b,g}
}{
\left\langle p_{\rm ref}\right\rangle_{b,g}
},
\\
&R_N
=\frac{\mathbb E[N_{\rm union}\mid N_{\rm ref}\geq1]}
{\mathbb E[N_{\rm ref}\mid N_{\rm ref}\geq1]},
\end{aligned}
\label{eq:countgain}
\end{equation}
where $\langle\cdot\rangle_{b,g}$ denotes the average over the uniform activity onset distribution and the lensing geometries with area weights $w_g$. 

\begin{figure*}[!htbp]
\centering\includegraphics[width=0.9\textwidth]{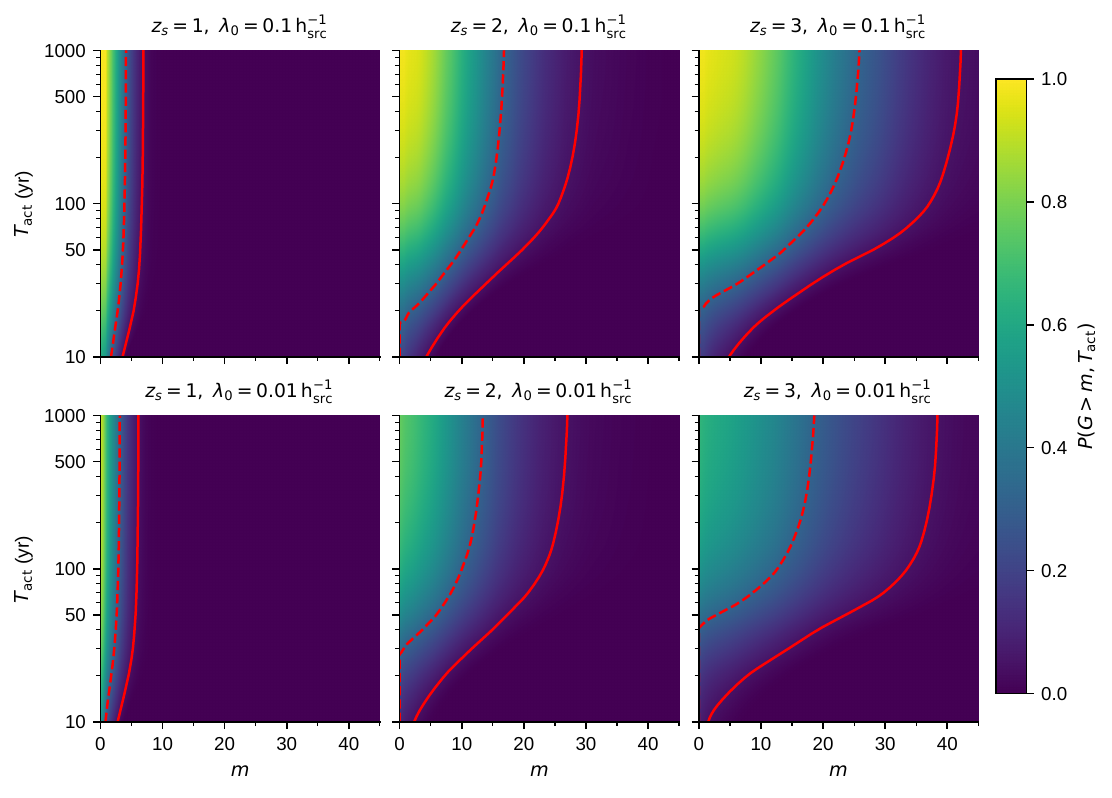}
\caption{Historical span exceedance probability $P(G>m,\Tact)$ for $m\in[0,45]$ and source-frame durations $\Tact\in[10,1000]$ yr. Columns correspond to $z_s=[1,2,3]$; while rows correspond to $\lambda_0=[0.1, 0.01]\,{\rm h}_{\rm src}^{-1}$, respectively. Red dashed and solid curves denote the $68\%$ and $95\%$ of $G$, respectively.}\label{fig:redshift}
\end{figure*}

\subsection{Joint Constraints on Activity-Span from Two Observable Signal Classes}\label{sec3-2}
For a intrinsic burst emitted at source-frame time $s$ with brightness $u$, let $Y_i(s,u)$ be the indicator that image $i$ is detected over the full monitoring duration as Eq.~(\ref{eqci}), and define $Y_{\rm tot}(s,u)=\sum_iY_i(s,u)$. For a detection-category indicator $f(s,u)$, the number of intrinsic bursts in that category is
\begin{equation}
\mu[f]=\nu_0\int_b^{b+T_{\rm act}} ds
\int_0^1 du\,f(s,u).
\end{equation}
The required dimensionless expected counts are
\begin{equation}
\begin{aligned}
&U=\mu[\mathbf{1}_{Y_{\rm tot}\geq1}],&
&B=\mu[\mathbf{1}_{Y_{\rm tot}\geq2}],\\
&V_1=\mu[Y_1\mathbf{1}_{Y_{\rm tot}=1}],&
&B_{2}=\mu[Y_1\mathbf{1}_{Y_{\rm tot}\geq2}],
\end{aligned}
\label{eq:categorymeans}
\end{equation}
where $U$ is the number of detected intrinsic bursts, $B$ is the number detected in at least two images, $V_1$ is the number of reference-only intrinsic bursts, and $B_{2}$ is the number of multi-image intrinsic burst that are also detected in the reference image. Under the Poisson burst process, the four disjoint categories correspond to independent Poisson counts with means $V_1$, $B_{2}$, $B-B_{2}$, and $U-V_1-B$, respectively: reference-only intrinsic bursts, multi-image intrinsic bursts including the reference image, multi-image intrinsic bursts excluding the reference image, and single-image intrinsic bursts excluding the reference image. To construct the joint count likelihood while accounting for the overlap between the two signal classes, we assign each source to one of four mutually exclusive outcomes and define the corresponding joint probabilities of outcome and sample inclusion. The four outcomes are: (i) all detected parents are reference-only, so no parent is detected in more than one image ($p_{00,\rm ref}$); (ii) at least one reference-only parent and at least one single-image parent in a non-reference stream are detected, but no multi-image parent (i.e., no echo) is detected ($p_{10,\rm ref}$); (iii) exactly one detected parent appears in the reference image and in at least one other image, with no other detected parents ($p_{01,\rm ref}$); and (iv) all other configurations satisfying $N_{\rm ref}\geq1$ ($p_{11,\rm ref}$). The joint probabilities of each outcome and sample inclusion are
\begin{equation}
\begin{aligned}
p_{00,{\rm ref}}&=e^{-(U-V_1)}(1-e^{-V_1}),\\
p_{01,{\rm ref}}&=B_{2} e^{-U},\\
p_{10,{\rm ref}}&=e^{-B}(1-e^{-V_1})(1-e^{-(U-V_1-B)}),\\
p_{11,{\rm ref}}&=p_{\rm ref}-p_{00,{\rm ref}}-p_{01,{\rm ref}}-p_{10,{\rm ref}}.
\end{aligned}
\label{eq:jointdiscovery}
\end{equation}
Within mass map $M_{\rm cl}$, average over its source-area weights and the onset distribution yields 
\begin{equation}
\pi_{c,M_{\rm cl}}=
\frac{\langle p_{c,{\rm ref}}\rangle_{b,g\mid M_{\rm cl}}}{\langle p_{\rm ref}\rangle_{b,g\mid M_{\rm cl}}},
\label{eq:classprob}
\end{equation}
where $c\in\{00, 01, 10, 11\}$. To preserve the historical reach of a detected cross-image association, we bin \(G\) using fixed edges at \(0\), \(1\), and \(10\)
\begin{equation}
\mathcal G_0=\{0\},\quad \mathcal G_1=(0,1],\quad
\mathcal G_2=(1,10],\quad \mathcal G_3=(10,\infty).
\end{equation}
A selected source therefore falls into one of eight mutually exclusive class-span outcomes. Their probabilities for mass map \(M_{\rm cl}\) are
\begin{equation}
\Pi_{ck,M_{\rm cl}}=
\frac{\left\langle p_{c,\rm ref}f_{k|c}\right\rangle_{b,g\mid M_{\rm cl}}}
{\langle p_{\rm ref}\rangle_{b,g\mid M_{\rm cl}}},
\label{eq:classgainprob}
\end{equation}
where $k$ labels the gain $G$ bin, $f_{k\mid c}\equiv P(G\in\mathcal G_k\mid c)$ denote the probability that the recovered historical gain lies in $\mathcal G_k$, conditional on at least one reference-image detection and observational class $c$. Summing over the span index \(k\) gives the class-marginal probability \(\pi_{c,M_{\rm cl}}\). 

For a hypothetical cohort of \(N\) selected, multiply imaged sources, let \(n_{ck}\) be the number of sources in class \(c\) and span bin \(k\), so that \(\sum_{c,k} n_{ck}=N\). Marginalizing over cluster mass map with prior weights $\alpha_{M_{\rm cl}}=\frac{\sum_{g\in M_{\rm cl}}w_g}{\sum_g w_g}$ yields the marginal likelihood of the outcome count vector $\boldsymbol n=\vec{n}_{ck}$
\begin{equation}
\mathcal L(\boldsymbol n)\propto
\sum_{M_{\rm cl}}\alpha_{M_{\rm cl}}\prod_{(c,k)}[\Pi_{ck,{M_{\rm cl}}}]^{n_{ck}}.
\label{eq:classlikelihood}
\end{equation}
Since different activity durations predict different fractions of sources with neither, only Class~I, only Class~II, or both associations, we compare hypotheses $A$ and $B$ using the likelihood ratio for a sample whose outcome-count vector is $\boldsymbol n$. We compare two activity duration hypotheses with the Bayes factor
\begin{equation}
B_{12}(\boldsymbol n)
=\frac{P(\boldsymbol n\mid H_1)}{P(\boldsymbol n\mid H_2)}=\frac{\mathcal L_1(\boldsymbol n)}{\mathcal L_2(\boldsymbol n)}.
\label{eq:bayesfactor}
\end{equation}
To quantify the ability of a repeating-FRB sample to distinguish between two activity durations, we compute the expected log Bayes factor under $H_1$
\begin{equation}
\begin{aligned}
\mathbb{E}[\ln B_{12}] =\sum_{\boldsymbol n:\,\sum_{c,k}n_{ck}=N}P(\boldsymbol n\mid H_1)\ln(B_{12}(\boldsymbol n).
\end{aligned}
\label{eq:expectedlogbayes}
\end{equation}
For a threshold of $\mathbb{E}[\ln B_{12}]=3$, the corresponding evidence ratio is approximately $20$, which we interpret as strong evidence for $H_1$ over $H_2$. 

\section{Results}\label{sec4}
\subsection{Historical Temporal Reach and Activity-Duration Constraints}
\begin{figure*}[!htbp]
\centering\includegraphics[width=0.9\textwidth]{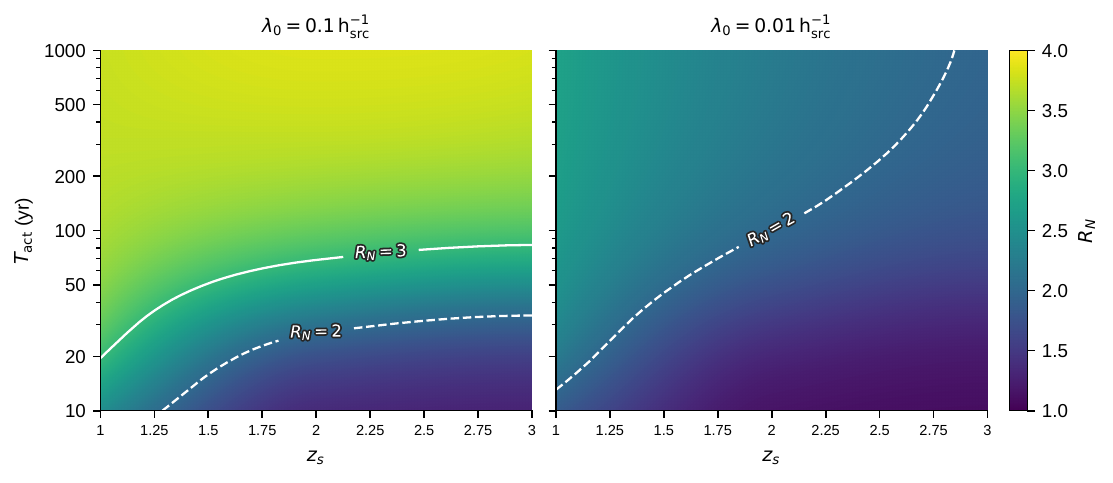}
\caption{Distinct intrinsic burst count gain $R_N$ as a function of source redshift $z_s\in[1,3]$ and source-frame activity duration $\Tact\in[10,1000]$ yr. The left and right panels correspond to $\lambda_0=0.1\,{\rm h}_{\rm src}^{-1}$ and $0.01\,{\rm h}_{\rm src}^{-1}$, respectively. White dashed and solid contours mark $R_N=2$ and $3$, respectively.}\label{fig:countsurface}
\end{figure*}

Galaxy-cluster lensing therefore not only extends the observable burst history of repeating FRBs within a finite monitoring campaign, but also provides additional burst records that can be used to study FRB environments and evolution. Figure~\ref{fig:redshift} shows the probability $P(G>m,\Tact)$ for activity durations $\Tact\in[10,1000]~{\rm yr}$, source redshifts $z_s=[1,2,3]$, and burst rates $\lambda_0=[0.1,0.01]\,{\rm h}_{\rm src}^{-1}$. In all cases, larger $\Tact$ increases the probability that distinct bursts are detected at source-frame epochs separated by a longer historical baseline, thereby enabling larger values of $G$. The probability nevertheless vanishes beyond a finite maximum gain threshold. For short activity episodes, this cutoff is set primarily by the activity duration, which limits the range of source frame epochs that distinct bursts can span. For long episodes, it is instead set primarily by the maximum historical span accessible through lensing. Setting $t_{ik}=T_{\rm obs}$ and $t_{jl}=0$ for an earlier image $i$ and a later image $j$ in Eq.~\eqref{eq:span}, and maximizing over images, yields the envelope
\begin{equation}
L_{\rm env}=\frac{\max_i\Delta_i-\min_i\Delta_i+T_{\rm obs}}{1+z_s}.
\label{eq:envelope}
\end{equation}
At $z_s=2$, the catalogue maximum is \GeometryCeiling\ source years. For episodes much longer than the sampled windows, increasing $\Tact$ only changes the probability of covering the onset and cannot extend the accessible history, so arbitrarily long episodes become indistinguishable. Finite exposure and missed detections further reduce the probability of satisfying the detection and association requirements at any given threshold, e.g., for $\Tact=100~{\rm yr}$ and $\lambda_0=0.1\,{\rm h}_{\rm src}^{-1}$, the 95\% gain threshold is $m_{95}\simeq25.6$, corresponding to a source-frame span $L_{95}\simeq42.6~{\rm yr}$. Comparing the upper and lower panels of Figure~\ref{fig:redshift} reveals how the burst rate affects the recovered historical span. Lensing time delays and the activity duration determine the accessible range of source epochs, whereas the burst rate controls how densely these epochs are sampled. A lower rate reduces the probability of detecting distinct bursts in widely separated source time windows, but it does not substantially alter the geometric or activity duration limits on $G$. The gain distributions in Figure~\ref{fig:redshift} extend to larger thresholds at higher source redshift for two reasons. First, the single-image source-frame baseline, $T_{\rm obs}/(1+z_s)$, becomes shorter, so the same recovered historical span corresponds to a larger gain in Eq.~\eqref{eq:gain}. Second, the higher-redshift lens models in our sample produce longer lensing delays and hence allow access to more widely separated emission epochs. A larger gain, however, need not correspond to a longer physical span, since $L_{\max}=G\,T_{\rm obs}/(1+z_s)$.

Beyond its gain in observable emission history, cluster lensing also increases the number of distinct intrinsic bursts detected. Figure~\ref{fig:countsurface} shows \(R_N\), the gain in the number of distinct intrinsic bursts detected when all lensed images are used instead of the reference image alone. For sources with at least one reference-image detection, \(R_N\) increases with the activity duration \(T_{\rm act}\) and gradually flattens at long activity timescales. This behavior arises because image time delays map the same observer frame interval to different source frame emission epochs: a longer activity episode allows the observing windows of delayed images to fall within the active episode more often, thereby contributing more distinct bursts. Once \(T_{\rm act}\) substantially exceeds the source frame span covered by these windows, further prolonging the activity yields diminishing additional gain. Higher burst rates produce larger gains, whereas low-rate samples are more strongly affected by the requirement of at least one reference-image detection. This selection relatively enhances the reference-image count, thereby suppressing the relative gain from additional bursts.

\begin{figure*}[!htbp]
\centering\includegraphics[width=0.9\textwidth]{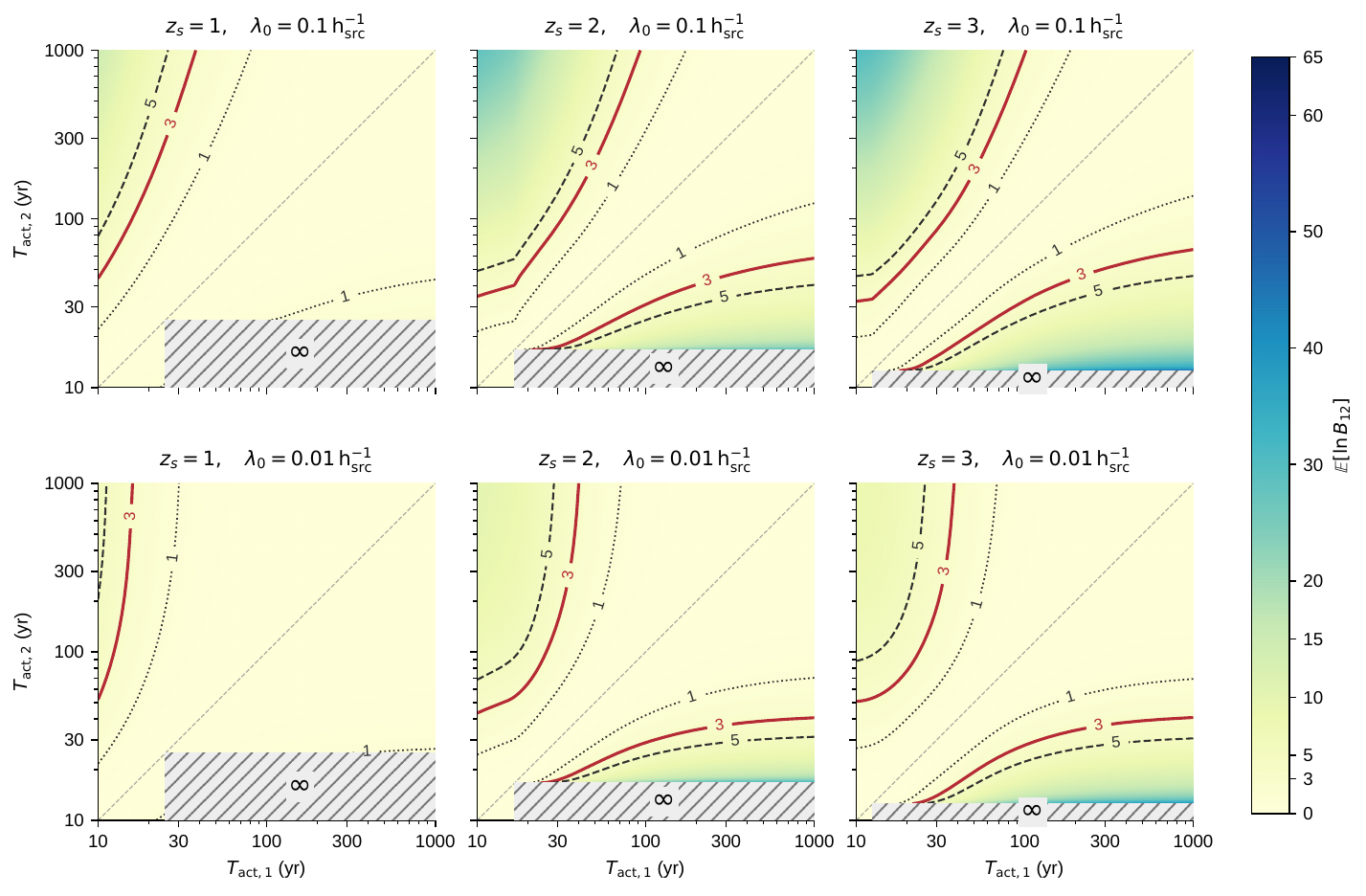}
\caption{Expected Bayes factors for $N=10$ repeating FRBs, combining the observational class and historical gain. $T_1$ and $T_2$ denote the generating and alternative activity durations, respectively. Columns correspond to $z_s=1,2,3$; while rows correspond to $\lambda_0=0.1,~0.01\,{\rm h}_{\rm src}^{-1}$. Colors show finite $\mathbb E[\ln B_{12}]$ without saturation, grey hatching marks $\mathbb E[\ln B_{12}]=+\infty$ from physical exclusions.}\label{fig:classification}
\end{figure*}

In addition to the gains in observable historical span and mean burst number, classification estimates based on different lensing signal types provide an additional way to distinguish between activity models. Figure~\ref{fig:classification} shows simulated strongly lensed repeating-FRB signals generated with activity duration $T_{\rm act,2}$ and analyzed with model $T_{\rm act,1}$, with the resulting lensing classifications used to compute Bayesian evidence following Section~\ref{sec3-2}. The strong discrimination arises when the two models disagree about whether a detected separation is physically possible ($\mathbb{E}[\ln B_{12}]>3$). A long activity episode can produce a span that a short episode forbids, which produces the $T_{\rm act,2}>T_{\rm act,1}$ region in Figure~\ref{fig:classification}. Such differences become appreciable only when the two durations differ by an order of magnitude. Conversely, a short episode cannot produce the signal expected from a long episode under any circumstances, corresponding to the grey $\ln \mathbb{E}[B_{12}]=+\infty$ region for $T_{\rm act,2}<T_{\rm act,1}$ in Figure~\ref{fig:classification}. This asymmetry motivates retaining both directions of the Bayes factor comparison. Different repeating-FRB scenarios predict distinct activity durations: young magnetars of decades to approximately a century~\citep{Metzger2017,MargalitMetzger2018}, young rotation-powered pulsars of tens to hundreds of years~\citep{Lyutikov2016}, binary-comb lifetimes of $\sim10^4$ yr~\citep{IokaZhang2020}, and accretion-powered phases ranging from short-lived unstable mass transfer lasting $\sim10$--$100$ orbital periods ($\sim3$--$30$ yr for a 100-day orbit) to much longer thermal-timescale transfer~\citep{Sridhar2021,SridharMetzger2022}. These differing predictions imply that the proposed test can potentially distinguish among these scenarios.

\subsection{Astrophysical Applications of Cluster-Lensed Repeating-FRB Activity Histories}

Cluster lensing historical spans not only constrain activity durations and thereby help distinguish among different repeating FRB scenarios, but also enable comparisons of the FRB environment on timescales otherwise inaccessible to current surveys. Dispersion measure (DM) traces the electron column, while rotation measure (RM) also depends on the line-of-sight magnetic field, so their evolution can test whether a changing propagation environment accompanies burst activity. A historical comparison is particularly useful for determining whether changes seen during recent monitoring persist over a much longer interval or represent a temporary disturbance. Both behaviors have observational motivation:~\citet{WangDecadal2025} report a long-term RM decline in FRB~20121102A, while~\citet{Ng2025} find diverse stochastic and secular behavior among repeaters. As a scale illustration, the approximately $7\times10^4\,{\rm rad\,m^{-2}}$ RM decrease observed for FRB~20121102A at $z=0.193$ would appear as about $10^4\,{\rm rad\,m^{-2}}$ at $z_s=2$ for the same intrinsic change, owing to the $(1+z_s)^{-2}$ dilution. It establishes a potentially relevant scale without extrapolating the measured trend to an unobserved epoch. The key observational ambiguity is that different images traverse different plasma paths, so their DM or RM difference can reflect source evolution, path-dependent propagation, or both \citep{Cordes2017,Er2020}. The two classes of lensing pairs play complementary roles, i.e., distinct burst pairs connect source epochs, whereas same burst pairs can help characterize propagation differences along the paths they share, provided the local contribution is common and the intervening screens are sufficiently stable. Combining these measurements could therefore distinguish a historical change in the source environment from a persistent difference between sight-lines.

An associated persistent radio source (PRS) provides a complementary probe of the engine's energy injection and its surrounding environment. If the emission originates from an engine-powered synchrotron nebula, its luminosity and spectrum reflect the accumulated particle population, magnetic field strength, and radiative and expansion losses, rather than the instantaneous FRB burst rate alone. Cluster lensing enables this comparison, i.e., images observed simultaneously sample the PRS at different source frame epochs, separated by $|\Delta_i-\Delta_j|/(1+z_s)$. For the example in Figure~\ref{fig:cluster}, the two more delayed images correspond to source states approximately 7.7 and 14.0 source years earlier than that in the earliest arriving image. After correcting for their effective magnifications, the image fluxes can therefore be compared to probe changes in PRS luminosity over these intervals, without requiring equally long monitoring. With the observed flux density given by $S_{\nu_{\rm obs}}=|\mu_{\rm PRS}|(1+z_s)L_{(1+z_s)\nu_{\rm obs}}/(4\pi D_L^2)$, a spectral luminosity of $10^{28}\,{\rm erg\,s^{-1}\,Hz^{-1}}$ at $z_s=2$ yields only $\sim1\,\mu{\rm Jy}$ even for $|\mu_{\rm PRS}|=10$, requiring deep radio observations.This luminosity lies within the range of known PRSs~\citep{Chatterjee2017,Bruni2024}. The evolution need not be a steady decline, e.g.,~\citet{Bhardwaj2025} find no significant seven-year L-band trend in the PRS of FRB~20121102A, whereas~\citet{Matsumiya2026} model an activity driven brightening followed by decades long fading for FRB~20240114A. Joint burst, RM, and PRS measurements can therefore test whether the inferred activity history and engine-nebula evolution are mutually consistent, once finite source size, scintillation, contamination, and magnification uncertainty are controlled before image differences are attributed to intrinsic evolution.

A third application of lensing time delays is to schedule multi-wavelength observations of later images once the first image of an event has been detected. The prediction of transient reappearances from lensing delays has precedents in supernova studies, including SN~Refsdal, whose reappearance was predicted and subsequently observed~\citep{Treu2016,Kelly2016}, and the predicted future lensed image of the likely Type~Ia SN~Requiem~\citep{Rodney2021}. A related strategy has been proposed for gravitational-wave early warning by \citet{Magare:2023hgs, Magare:2026ssh}, i.e., lensing is used to predict and target a later image of an already detected event. In the present application, the same logic is applied to multi-wavelength follow up of lensed repeating FRB images. Combined with archival searches for other bursts from the same source that sample earlier source epochs, this converts a serendipitous detection into a targeted follow-up search with a delay-dependent observing window. The main limitation of such a search is sensitivity: for a co-spatial counterpart burst with isotropic-equivalent energy \(E_{\rm iso}\), the bolometric fluence is \(\mathcal F=|\mu|(1+z_s)E_{\rm iso}/(4\pi D_L^2)\); for a finite observed band, the same expression applies with the band-integrated emitted energy and the appropriate \(k\)-correction. At \(z_s=2\) and \(|\mu|=10\), an energy of \(10^{40}\) erg, comparable to the hard X-ray burst associated with the Galactic magnetar FRB~\citep{Tavani2021}, yields only about \(10^{-17}\,{\rm erg\,cm^{-2}}\). A detection would therefore constrain both the counterpart energy and its arrival time relative to the radio burst. Even without a detection, sufficiently deep observations covering the predicted arrival window would place upper limits on the high-energy emission, thereby testing models that predict brighter counterparts.

\section{Conclusions}\label{sec5}
The active lifetimes of repeating-FRB engines remain poorly constrained because short campaigns probe only a limited range of source-frame epochs. Galaxy-cluster lensing alleviates this limitation by allowing a single campaign to sample emission epochs separated by decades, thereby turning the delay-corrected separation of distinct bursts into a direct constraint on the activity span. Based on the cluster-lens model of MACS~J0717.5+3745, we quantify this historical reach under finite activity episodes, observing windows, and discovery selection. For a source-frame episode of duration $100~{\rm yr}$ at $z_s=2$ and a source-frame rate of $0.1\,{\rm h}_{\rm src}^{-1}$, a five-year campaign conditioned on discovery yields a $95\%$ gain threshold $m_{95}\simeq25.6$, corresponding to a source-frame span $L_{95}\simeq42.6~{\rm yr}$. At the population level, combining distinct-burst associations with same-burst echoes can test specified activity durations with a finite lens catalogue, distinguishing durations that differ by an order of magnitude at an expected Bayes factor exceeding $20$. More broadly, these lensed source-frame epochs enable applications beyond activity-duration tests by linking burst activity to the evolution of the engine and its environment: the burst span constrains the epochs of activity; propagation observables probe the magnetized plasma along the sight lines; persistent emission traces energy injection and losses; and multiwavelength counterparts constrain the energy release and emission processes of individual bursts. Cluster lensing enables these tests by providing access to earlier source epochs and, when delays are constrained, by guiding observations of later-arriving images.

Our results should be interpreted with several caveats. First, the recovered burst span constrains when the source was capable of producing detectable bursts, not whether it remained continuously active. Two widely separated bursts cannot distinguish sustained activity from recurrent episodes separated by quiescence, and a non-detection at an intervening epoch may reflect changes in luminosity, spectrum, or the surrounding plasma rather than an inactive engine. A physically important extension is therefore to connect the reconstructed burst history with the evolution of energy supply and radio visibility by comparing burst properties, DM, RM, and persistent emission across lens-separated epochs, provided that propagation differences between image paths are constrained. Second, our calculations are based on a single cluster rather than a representative lens population. Extending the analysis to other clusters would identify systems that provide not only long delays but also usable historical baselines supported by detectable images, reliable delay constraints, and feasible monitoring. These extensions would broaden the present activity-span test into a probe of how FRB engines sustain, interrupt, or renew their activity.

\begin{acknowledgements}
This work is supported by the National Natural Science Foundation of China under Grants Nos.\ 12603003, 12322301, 12275021, and 12503002; the National Key R\&D Program of China under Grant No.\ 2024YFC2207400; the Science and Technology Research Project of Hubei Provincial Department of Education under Grant No.\ Q20251302; and the Postdoctoral Project of Hubei Province under Grant No.\ 2026HBBSHXF091
\end{acknowledgements}

\paragraph{Data availability.}
The CATS v4.1 lens models of MACS~J0717.5+3745 used in this work are publicly available through the STScI Hubble Frontier Fields archive~\citep{HFFArchive}.

\bibliographystyle{aa}
\bibliography{references}
\end{document}